\documentclass{iau} 
\usepackage{graphicx}
\usepackage{ctable}
\usepackage{amsmath}	
\usepackage{amssymb}	
\usepackage{float}
\usepackage{natbib}

\newcommand\aap{A\&A}  
 
\newcommand\apj{ApJ}  
\newcommand\apjl{ApJ}           
     
\newcommand\mnras{MNRAS}
\newcommand\nat{Nature}

\title[Star-planet interaction \& spectral lines] 
{Star-planet interaction through spectral lines}

\author[Villarreal D'Angelo et al.]  
{C. Villarreal D'Angelo$^1$, A. A. Vidotto$^1$, A. Esquivel$^{2,3}$, M. A. Sgr\'o$^3$, T. Koskinen$^4$  \and L. Fossati$^5$  }

\affiliation{$^1$School of Physics, Trinity College Dublin, The University of Dublin \\ College Green, Dublin 2,
Dublin, Ireland. email: {\tt villarrc@tcd.ie} \\
$^2$Instituto de Ciencias Nucleares, UNAM, Ciudad de M\'exico, M\'exico\\
$^{3}$Instituto de Astronom{\'i}a Te{\'orica} y Experimental, Conicet-UNC, C{\'o}rdoba, Argentina
$^{4}$University of Arizona, Lunar and Planetary Laboratory, Tucson, Arizona, United States\\
$^{5}$ Space Research Institute, Austrian Academy of Sciences, Graz, Austria
}

\pubyear{2020}
\volume{354}  
\jname{Solar and Stellar Magnetic Fields: Origins and Manifestations }
\editors{Kosovichev, Strassmeier \& Jardine, eds.}
\begin{document}

\maketitle

\begin{abstract}
The growth of spectroscopic observations of exoplanetary systems allows the possibility of testing theoretical models and studying the interaction that exoplanetary atmospheres have with the wind and the energetic photons from the star. In this work, we present a set of numerical 3D simulations of HD 209458b for which spectral lines observations of their evaporative atmosphere are available. The different simulations aim to reproduce different scenarios for the star-planet interaction. 
With our models, we reconstruct the Ly$\alpha$ line during transit and compare with observations. The results allows us to analyse the shape of the line profile under these different scenarios and the comparison with the observations suggest that HD209458b may have a magnetic field off less than 1 G. We also explore the behaviour of the magnesium lines for models with and without magnetic fields.

\keywords{stars: individual, HD209458, stars: planetary systems, stars: winds, outflows, line: profiles, magnetic fields, methods: numerical}
\end{abstract}

\firstsection 

\section{Introduction}
The Lyman $\alpha$ line has been broadly used to detect escaping atmospheres of close-in exoplanets. The first example was the case of HD209458b, a hot-Jupiter orbiting a solar-like star.  While observed in transit, this exoplanet revealed an absorption of 15$\%$,  ten times larger than the one produced at optical wavelengths \citep{Vidal-Madjar2003}. This implied that the obstacle obscuring the disc of the star was bigger than the size of the planet Roche lobe. The Lyman $\alpha$ observation also revealed that the duration of the transit in this line was longer than the one produced by the opaque planetary disc, probing the existence of a cometary tail of neutral hydrogen trailing the planet. Finally, the absorption at high blue-shifted velocities ($\sim$100 km s$^{-1}$) indicated that neutral hydrogen atoms were accelerated away from the star.  

In transit observations in other lines like O$_\mathrm{I}$, C$_\mathrm{II}$, Si$_\mathrm{III}$ and Mg$_\mathrm{I}$ \citep{Vidal-Madjar2004, Linsky2010, Vidal-Madjar2013, Bourrier2014} also presented an excess of absorption compared to the optic, confirming the blow-off state of this atmosphere and showing that lighter elements were dragging heavier elements towards the upper atmosphere. But only lighter atoms like hydrogen have been found so far above the Roche lobe. 

Many theoretical works since then have been developed in order to reproduce the  transit observations of HD209458b in Lyman $\alpha$. In particular, 3D numerical simulations that include the stellar wind and the escaping planetary atmosphere with or without magnetic fields \citep{Villarreal2014, Villarreal2018}, and with or without charge exchange \citep{Esquivel2019}, have been able to reproduce the observations. 3D simulations can capture the asymmetric nature of the interaction   between the stellar and planetary wind which may leave an imprint in the line profile observed during transit.

\section{Numerical simulations}
A number of numerical simulations have been run using the MHD code Guacho\footnote {Available for free at https://github.com/esquivas/guacho}. The physical domain of the simulations comprises the star at the centre and the planet orbiting around it. The models focused on the system HD209458 and assume that the planet has a wind as a consequence of the heating of the planetary atmosphere.     
We only model hydrogen (ionised for the stellar wind and partially ionised for the planetary wind) but other elements can be accounted assuming an abundance relative to hydrogen (like magnesium). The simulations account for the gravity of the star and the planet, the radiation pressure and the ionising flux from the star.  In addition, a radiative transfer module takes care of the change in the density of neutral hydrogen due to collisional ionisation, photoionisation and recombination. Heating due to photoionisation and cooling due to the atomic transitions of hydrogen are included. More detailed information of the model setup and the code itself can be found in \cite{Villarreal2018}.

The stellar wind is launched from the stellar surface with the initial conditions set by a thermally driven wind with a coronal temperature of 1.5 MK. The planetary wind is launched from 3R$_p$ with initial conditions set by the 1D atmospheric model of \cite{Murray-Clay2009}. System parameters used in the models together with the winds values are presented in Table \ref{tab:1}. We consider that both objects, star and planet, posses a dipolar magnetic field perpendicular to the orbital plane and both aligned.  We ran a total of five simulations varying the strength of the magnetic field both in the star and the planet. Table \ref{tab:2} summarises the model names and the corresponding values of B at the stellar and planetary poles.

\ctable[
	caption= {Stellar and planetary winds parameters employed in the simulations for the system HD 2019458. System parameters taken from \cite{Torres2008}. (1)From \cite{Sanz-Forcada2011}.},
    label= {tab:1}
]{llc}{
}{
	\toprule
    \toprule
    {\bf Stellar parameters}                 & Sym.                 & HD 2019458\\
    \midrule
    Radius $[R_\odot]$                       & $R_{\star}$          & $1.2$\\
    Mass $[M_\odot]$                         & $M_{\star}$          & $1.1$\\
    Wind temperature [MK]                    & $T_\mathrm{\star}$   & $1.5$\\
    Mass loss rate $[M_{\odot}~\mathrm{yr}^{-1}]$     & $\dot{M_{\star}}$    & $2.0\times10^{-14}$\\
    Photon rate $[$s$^{-1}]$                   & $S_0$                & $2.5\times 10^{38}$ $^{(1)}$\\
    \midrule
    \midrule
    {\bf Planetary  parameters}              & Sym.                 & HD 209458b\\
    \midrule
    Radius $[R_\mathrm{J}]$                         & $R_\mathrm{p}$       & $1.38$\\
    Mass $[M_\mathrm{J}]$                           & $M_\mathrm{p}$       & $0.67$\\
    Orbital period [d]                       & $\tau_\mathrm{p}$       & $3.52$\\
    Inclination [deg]                    & $i$                  & $86.71$\\
    Wind launch radius $[R_\mathrm{p}]$               & $R_\mathrm{w,p}$     & $3$\\
    Wind velocity at $R_\mathrm{w,p}$ [km s$^{-1}$] & $v_\mathrm{p}$       & $10$\\
    Wind temperature at $R_\mathrm{w,p}$ [K]        & $T_\mathrm{p}$       & $1 \times10^4$\\
    Ionisation fraction at $R_\mathrm{w,p}$         & $\chi_\mathrm{p}$                & 0.8\\
    Mass loss rate [g s$^{-1}]$              & $\dot{M}_\mathrm{p}$ & $2 \times 10^{10}$\\
    \bottomrule
    \bottomrule
}

\ctable[
	caption={Models characteristics and the estimated values for the stand-off distance at the sub-stellar point. Last column shows the integrated Ly$\alpha$ absorption in the velocity range of $\pm 300$ km s$^{-1}$.},
    label= {tab:2},
]{lcccc}{
}{
 \toprule
    {\bf Models} & $B_{\star}$ [G]	& $B_\mathrm{p}$ [G]	& $R_0$ [$R_\mathrm{p}$] & (1-I/I$_\star$) [$\%$]\\
    \midrule
    B1.0	 & 1				& 0			& 7             & 12.1 \\
    B1.1	 & 1				& 1			& 7             & 12.1 \\
    B1.5	 & 1				& 5			& 9             & 10.7 \\
    B5.1	 & 5				& 1			& 4             & 12.7 \\
    B5.5	 & 5 			    & 5			& 4             & 13.4 \\
    \bottomrule
}

\subsection{Lyman $\alpha$ line under different magnetic field values}
\begin{figure}
\begin{center}
\includegraphics[width=\textwidth]{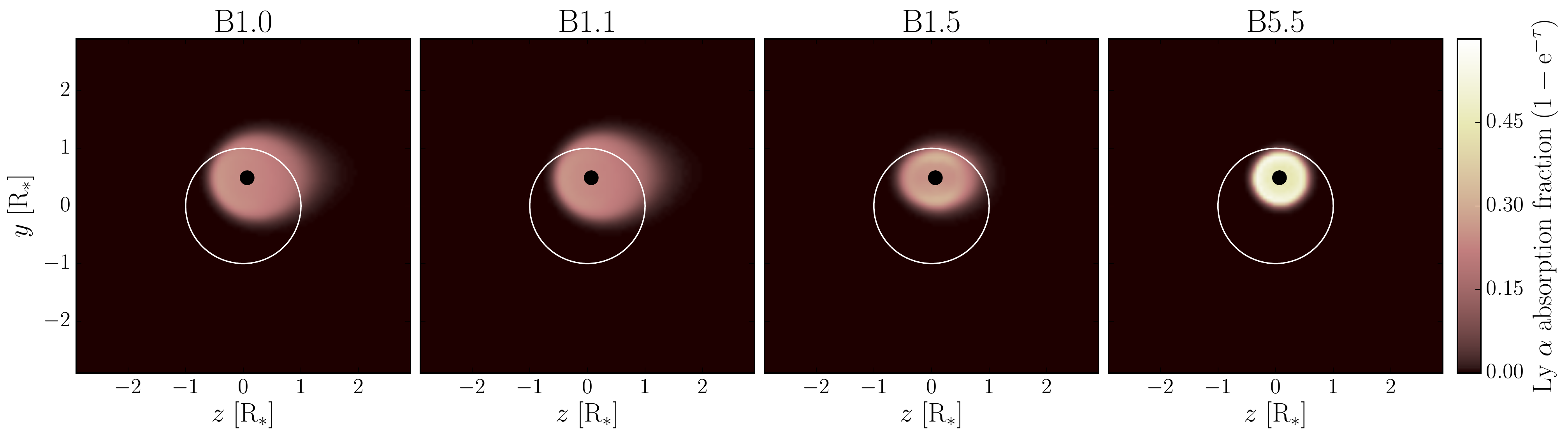} 
 \caption{Lyman $\alpha$ absorption fraction integrated between [-300, 300] km s$^{-1}$, for four of the models presented in Table \ref{tab:2}. The plot is a zoom at the position of the star (white circle) at the moment of mid-transit. The planet is represented with the black filled circle.}
   \label{fig1}
\end{center}
\end{figure}

For each model we calculated the Ly$\alpha$ absorption fraction (I/I$_\star=(1-e^{-\tau})$) in the line of sight (LOS) direction integrated over the velocity range  [-300, 300] km s$^{-1}$ and taking into account the orbital inclination. The optical depth is computed assuming a Voigt profile with an absorption cross section $\sigma$=0.01105 cm$^2$. The results are shown in Figure \ref{fig1}. From the figure we can see that models with a stellar magnetic of 1 G  (B1.0, B1.1 and B1.5) presents a more extended area of neutral hydrogen around the planet than models with a higher stellar magnetic field (B5.1 and B5.5). Since a higher B$_\star$ increase the velocity and temperature of the stellar wind, it confines the escaping planetary material to a smaller and spherical region. Figure \ref{fig1} only shows model B5.5 as model B5.1 gave similar results.  

For the models with B$_\star$=1 G, the planetary material stop the stellar wind further from the planet and the neutral hydrogen expands producing a more extended region. In this scenario, models with B$_p$=0 or 1 G behave similarly, indicating that the gas pressure in the atmosphere dominates over the magnetic pressure. Model B1.5 however, does show the influence of the planetary magnetic field as a higher  B$_p$ prevents the escape of material from the equatorial region (neutral are coupled to the ions) and ionise the polar one,  and so the region of neutral material around the planet have a more oval distribution.  

\begin{figure}
\begin{center}
\includegraphics[width=0.5\textwidth]{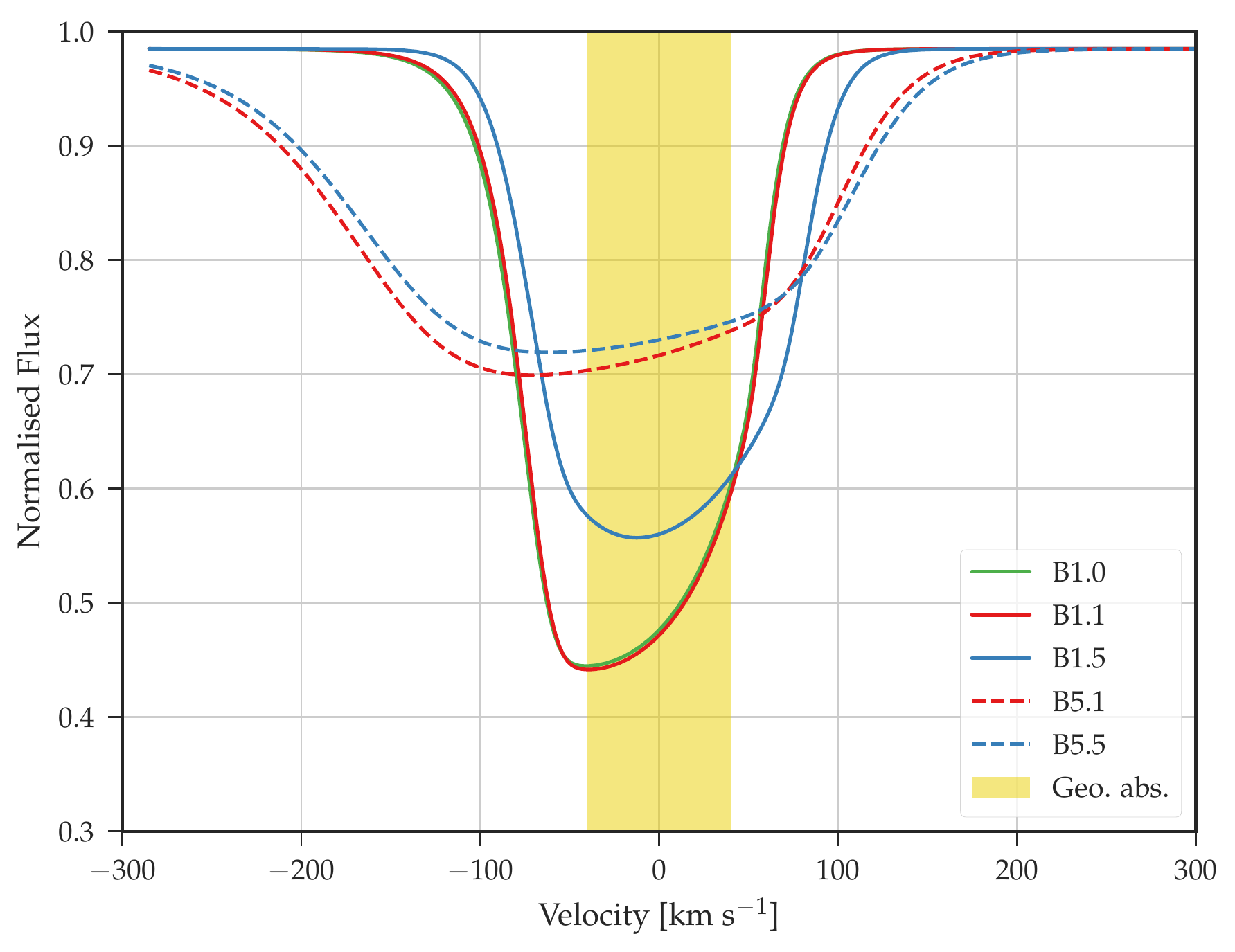} 
 \caption{Normalised stellar transmission as a function of the LOS velocity in the Ly$\alpha$ line averaged over the stellar disk as seen by an observer. The yellow stripe corresponds to part of the line contaminated with the geo-coronal glow. Solid lines represent models with B$_{\star}$=1 G and dashed lines for B$_{\star}$=5 G. Line colours represent different values for the planetary magnetic field. }
   \label{fig2}
\end{center}
\end{figure}

Figure \ref{fig2} shows the normalised transmission spectra in the Ly$\alpha$ line as a function of the LOS velocity integrated over the stellar disc. Models with B$_\star$=1 and 5 G are represented with solid and dashed lines respectively and the different B$_p$ values with colours. 

As mentioned before, models with a higher stellar or planetary magnetic field differentiate from the ones with a lower B (stellar or planetary) due to the shape of the neutral region around the planet, producing different depths  and shapes of the line profile. 

Whit a stronger stellar wind (B$_\star$=5 G) most of the neutral material is confined very close to the planet, having the smaller standoff distance at the sub-stellar point (see Table \ref{tab:2}). However, a small amount still escape away from the planet forming a small tail due to the interaction. The line shape is then asymmetric but with a lower depth. 

Another asymmetric line shape is obtained for models B1.0 and B1.1. In these cases, the stellar wind is not as strong (B$_\star$=1 G) and the neutral material is able to expand more, forming a larger cometary tail. In these cases, the interaction with the stellar wind doesn't accelerate the neutrals to a very high velocity but the column density is larger producing a larger absorption depth. 

A symmetric line shape is possible when B$_\star$=1 G and B$_p$=5 G (model B1.5). In this case, the planetary magnetic field constrain the expansion of the neutral material into a more spherical region at the equator. The velocities of the neutral material seems to be more constrained by the planetary magnetic field than the stellar wind.

Overall, the line shape for each model reveal that the interaction with the stellar wind can accelerate the neutral material to high blue-shifted velocities, and that these velocities are related to the type of interaction produced. \\

{\bf Comparison with observations:} Among the Ly$\alpha$ lines produced with the magnetic models presented above we can see that only models with B$_p$=0 or 1 G agreed with the 10$\%$ of absorption spotted at velocities of -100 km s$^{-1}$ in the observed line profile. 

The total absorption value is calculated integrating the line in Fig. \ref{fig2} over  [-300, 300] km s$^{-1}$ excluding the geocoronal emission range from [-40,40] km s$^{-1}$. The computed values are shown in Table \ref{tab:2} for all the models together with the model's stand-off distance. Most of them agree within the errors with the 15$\%$ of absorption found in the same velocity range by \cite{Vidal-Madjar2003}, although the lowest total absorption corresponds to the model with the largest stand-off distance (model B1.5).   

\begin{figure}
\begin{center}
\includegraphics[width=0.5\textwidth]{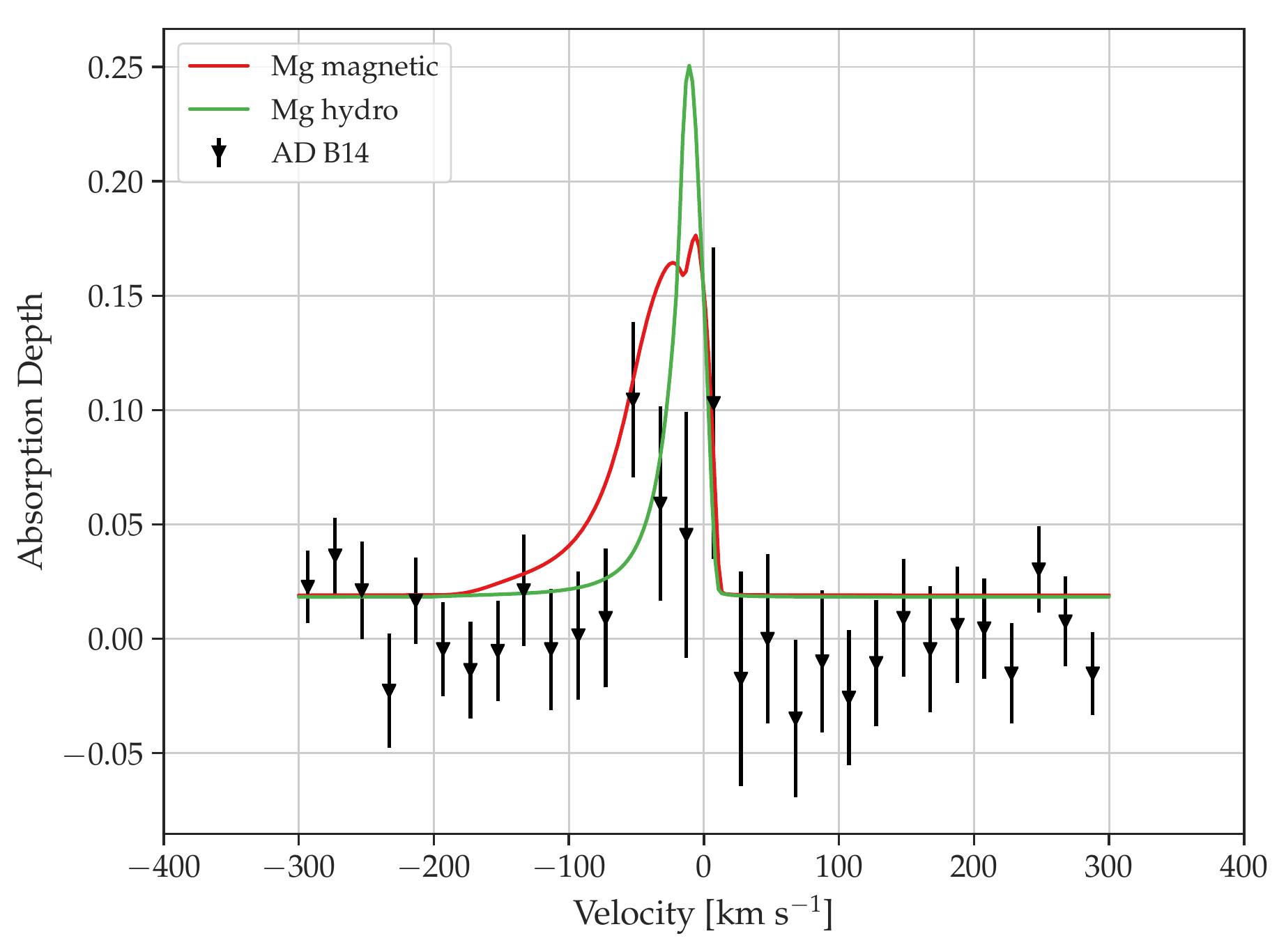} 
\includegraphics[width=0.49\textwidth]{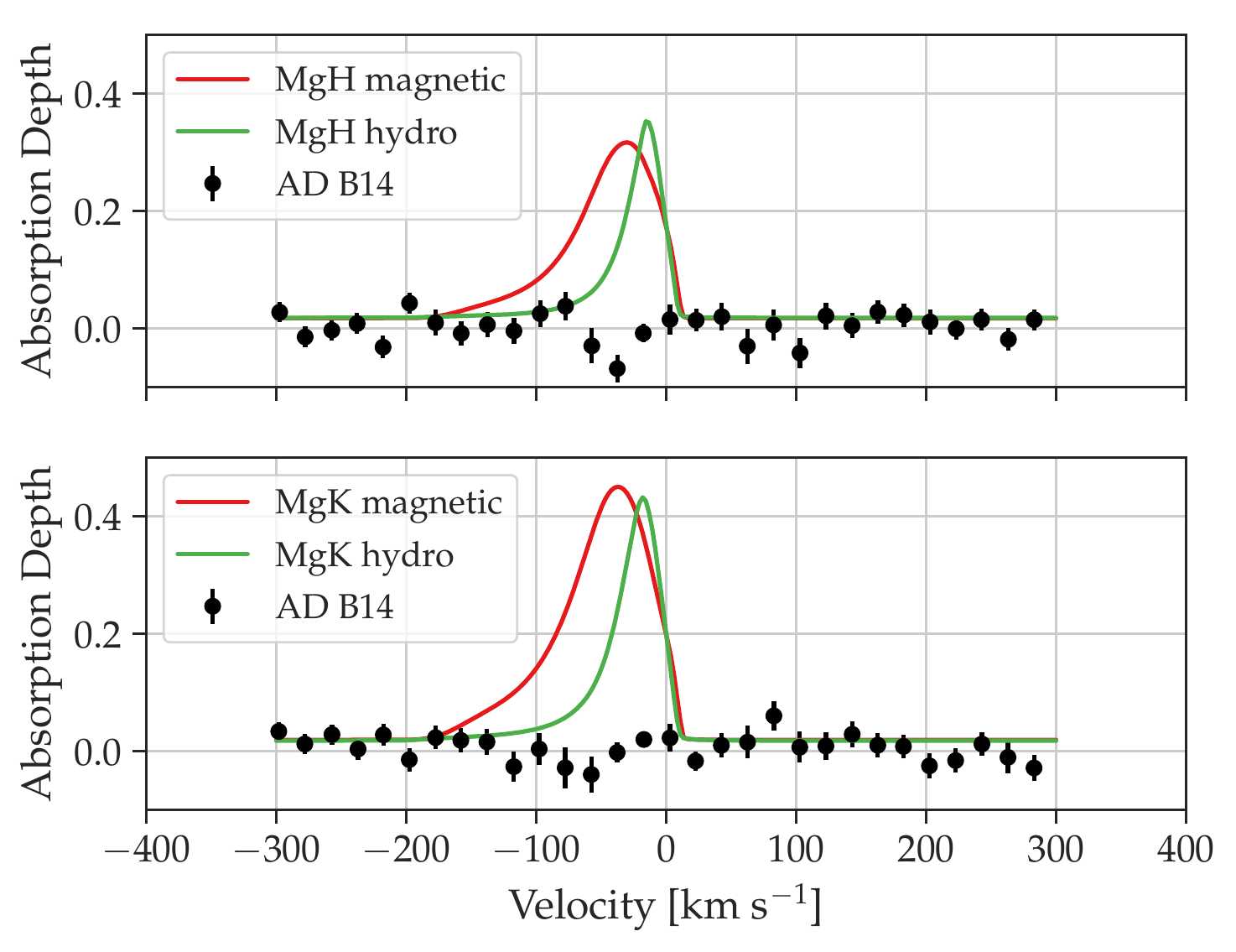} 
 \caption{Magnesium absorption fraction as a function of the LOS velocity for two models: hydrodynamic (green) and magnetohydrodynamic (red) with the similar initial conditions for planetary and stellar wind. Left: Mg$_\mathrm{I}$. Right top: Mg$_\mathrm{II}$ H. Right bottom: Mg$_\mathrm{II}$ K. Observational values presented in \cite{Bourrier2014} are shown with black dots.}
   \label{fig3}
\end{center}
\end{figure}

\subsection{The Mg lines}
To explore different type of interactions in other spectroscopic lines we have used one of the magnetic models presented above (model B1.1) together with a pure hydrodynamical model, with similar initial conditions for the stellar and planetary wind, from the work of \cite{Esquivel2019} (model M5a). For these two models we compute the abundance and calculate the absorption profile during transit for the Mg$_\mathrm{I}$ and Mg$_\mathrm{II}$ H $\&$ K lines. This was done as a post-processing from the output of those simulations. The results shown in this section are preliminary and so, they should be taken as such. The final purpose of these calculations is to compare the shape of the magnesium lines with two different sets of simulations (with and without magnetic fields). 

To calculate the abundance of magnesium within the simulations we used a 1D hydrodynamic escape model from T. Koskinen (private comm.) that includes several species (hydrogen, magnesium, helium, sodium). As these models predict a higher mass loss rate for HD209458b than the one we used in our models, we scaled the total hydrogen in our simulations to match the total hydrogen in the 1D model at the position of our boundary for the planetary wind. The relative abundance of magnesium (neutral and ionised) to total hydrogen computed in for the 1D model is fairly constant above 1.5R$_p$ and so we used this ratio as a scale factor to convert from hydrogen to magnesium in our 3D models. Below 3R$_p$ the magnesium abundance  follows the one predicted in the 1D model.
An important thing to have in mind with these calculations is that we are assuming that magnesium will follow hydrogen above the 3R$_p$ and have the same photoionisation rate which may not be the case. 

In the same way as for Ly$\alpha$, we calculated the absorption fraction of Mg$_\mathrm{I}$ and Mg$_\mathrm{II}$ H $\&$ K in both models with cross-section coefficients of $\sigma_\mathrm{Mg_I}=0.04869$ and $\sigma_\mathrm{Mg_{II}}=0.00814$ cm$^2$ respectively. The lines profiles are shown in Figure \ref{fig3} together with the observational results from \cite{Bourrier2014}. It is clear that the model including magnetic fields produce a broader line shape compared with the hydrodynamic one with absorption present a high blue-shifted velocities. However, both of them shown a sharp jump in the red part of the line. Also, both models predict absorption in the magnesium II H $\&$ K lines which is not found in the observations.

\section{Conclusion}
We have shown how the spectral lines such as Ly$\alpha$ and the Mg lines are shaped trough the interaction of the stellar and planetary winds. We have used the output of numerical simulations based on the HD209458 system to compute the transmission spectra in these lines during transit. We have shown that the Lyman $\alpha$ profile is shaped by magnetic fields and that a strong stellar magnetic field or a strong planetary magnetic field (5 G) with a weak stellar magnetic field (1 G) gives symmetric profiles. 
From these magnetic models we have calculated the total Lyman $\alpha$ absorption between [10-13]$\%$ matching the amount computed from the observations. 

The future launch of the Colorado Ultraviolet Transit Experiment (CUTE) satellite will produce a number of in-transit spectra in the NUV. Most prominent lines like magnesium will then be sampled for a number of extrasolar close-in gaseous planets. Our attempt was to study the behaviour of these lines in a very simplistic way assuming that magnesium will follow hydrogen above the 3R$_p$. 
The lines shape found in these cases show lines that are less broad for the hydrodynamic models than magnetic models. 

Spectral lines observed during a planetary transit can probe the escaping atmosphere of the planet and give an insight of the type of interaction that occurs between the planet and the star. We have shown that the presence of absorption in higher blue-shifted velocities indicates a strong interaction with the stellar wind. With the possibility to observe simultaneously several spectral lines in the future, for example with CUTE, and with the use of numerical models capable to reproduce such observations, we will be able to constrain the physical parameters of a planetary system.

\section*{Acknowledgements}
I want to thank the organisers to allow me to give this talk in this very rich symposium and for such a good organisation and predisposition to help. 
I also want to acknowledge the funding from the Irish Research Council through the postdoctoral fellowship. Project ID: GOIPD/2018/659.

\begin{discussion}

\discuss{L. Campusano}{Hydrodynamic vs magnetohydrodynamic models. Can one of them be discarded by observations?}

\discuss{C. Villarreal D'Angelo}{Not really as there is not too many observations and all of them have large errors too. Also models are dependent on initial conditions.}

\discuss{W. Cauley}{Can you guess what is causing the Mg$_\mathrm{II}$ models to be off?}

\discuss{C. Villarreal D'Angelo}{Not at this point.} 
\end{discussion}

\end{document}